\documentclass[prl,twocolumn,showpacs,superscriptaddress,preprintnumbers,amssymb]{revtex4-2}
\usepackage{times}
\usepackage{CJK}
\usepackage{fontsize}
\changefontsize{10}
\usepackage{graphicx}
\usepackage{float}
\usepackage{stmaryrd}
\usepackage[dvipsnames]{xcolor}
\usepackage{dcolumn}
\usepackage{bm}
\usepackage{dsfont}

\usepackage[percent]{overpic}

\usepackage{physics}
\usepackage[caption=false]{subfig}
\usepackage{comment}
\usepackage{manfnt}
\usepackage{svg}
\usepackage{placeins}
\usepackage{flafter}
\usepackage[colorlinks,allcolors=Blue,linktocpage]{hyperref}
\usepackage{soul}
\usepackage{multirow}
\usepackage{tikz}
\usetikzlibrary{calc,arrows.meta,angles,quotes}

\definecolor{eyepink}{RGB}{220,72,125}

\usepackage[bottom]{footmisc}
\usepackage{amsfonts}
\usepackage{wasysym}
\usepackage{ifsym}
\usepackage[export]{adjustbox}
\usepackage{dsfont}
\usepackage{cleveref}
\usepackage{lipsum}
\usepackage{stmaryrd}
\usepackage{MorrisIn,lettrine,Romantik}

\usepackage{cancel}
\usepackage{booktabs}
\usepackage{array}
\usepackage{appendix}

\newcommand{\rrangle}{\rangle \! \rangle}

\def\be{\begin{equation}}
\def\ee{\end{equation}}
\def\bea{\begin{eqnarray}}
\def\eea{\end{eqnarray}}

\newcommand{\id}{\mathds{1}}

\makeatletter\@addtoreset{equation}{section}\makeatother

\def\be{\begin{equation}}
\def\ee{\end{equation}}
\def\bea{\begin{eqnarray}}
\def\eea{\end{eqnarray}}
\def\ie{\begin{equation}\begin{aligned}}
\def\fe{\end{aligned}\end{equation}}

\newcommand{\A}{{\alpha}}

\newcommand{\D}{{\delta}}

\newcommand{\cE}{{\mathcal E}}

\newcommand{\cH}{{\mathcal H}}

\newcommand{\cN}{{\mathcal N}}
\newcommand{\cO}{{\mathcal O}}

\newcommand{\cT}{{\mathcal T}}

\newcommand{\mZ}{{\mathbb Z}}
\newcommand{\mR}{{\mathbb R}}

\newcommand{\mf}{\mathfrak }
 
\newcommand{\la}{\langle}
\newcommand{\ra}{\rangle}

\begin{document}

\setlength{\skip\footins}{0.5cm}
\addtolength{\textheight}{0.155in}  
\addtolength{\topmargin}{-0.06in}

\title{Eye-opening bounds on cusps}
\author{Ryan A. Lanzetta}
\email{rlanzetta@pitp.ca}
\affiliation{Perimeter Institute for Theoretical Physics, Waterloo, Ontario N2L 2Y5, Canada }
\author{Ian Moult}
\affiliation{Department of Physics, Yale University, New Haven, CT 06511, U.S.A.}
\author{Yifan Wang}
\affiliation{Center for Cosmology and Particle Physics, New York University,
New York, NY 10003, U.S.A.}
\affiliation{Institute for Advanced Study, Princeton, NJ 08540, USA}

\begin{abstract}
We derive new nonperturbative inequalities on the cusp anomalous dimensions of conformal line defects. We impose reflection positivity, locality, and conformal invariance on a pair of cusps forming an ``eye'' geometry, deriving a novel condition on the cusp anomalous dimension which we refer to as conformal concavity. The resulting constraint is much stronger than the known angular concavity, and our derivation applies also to cusps involving distinct line defects. Remarkably, it directly links the smooth and fusion limits, yielding bounds on defect-changing operator dimensions, Casimir energies, and subleading fusion data. We thus uncover a new quantitative bridge between aspects of the local operator data and the fusion rules of line operators in conformal field theories.
\end{abstract}

\maketitle
\lettrine[lines=3]{S}{ingularities} are powerful probes of Quantum Field Theory (QFT), whose constrained short-distance structure isolates basic universal data of the theory. They become particularly rich in the context of cusped line defects, which describe, among other things, heavy particles traveling in medium subject to a sudden kick. 
For heavy charged particles described by Wilson lines in gauge theories, the original setting in which cusps were studied \cite{Polyakov:1980ca,Dotsenko:1979wb,Brandt:1981kf,Korchemsky:1985xj,Korchemsky:1987wg,Isgur:1989vq,Korchemsky:1991zp}, an abrupt change in velocity produces radiation across a broad range of scales, thus encoding rich physics.
\par 
As point-like objects, cusps require point-like renormalization governed by the cusp anomalous dimension $\Gamma_{a \bar a}(\theta)$ \footnote{In our companion work \cite{exciting_magic}, we denote the cusp anomalous dimension by $\Gamma^{ab}(\theta)$. In each respective work these choices are more convenient for the presentation.}, where $\pi-\theta$ is the deflection angle and $a$, $\bar a$ refer to the line defect and its conjugate. The same quantity controls universal infrared logarithms in form factors and scattering amplitudes \cite{Korchemsky:1985xj,Korchemsky:1987wg,Korchemsky:1991zp}. 
In the lightlike limit these logarithms are enhanced to double logarithms, producing Sudakov suppression
\cite{Sudakov:1954sw,Korchemsky:1988hd}
and organizing much of the universal infrared structure of gauge-theory amplitudes
\cite{Korchemskaya:1992je,Catani:1998bh,Bauer:2000yr,Becher:2009cu,Gardi:2009qi}.
Cusps also arise in condensed matter systems near criticality, controlling universal finite-size contributions to entanglement and R\'enyi entropies, disorder operators, and other extended observables \cite{Casini:2006hu,Inglis:2013rrh,Stoudenmire:2014hja,Kallin:2014oka,Bueno:2015qya,Bianchi:2015liz,Witczak-Krempa:2016jhc,Wang:2021lmb,Wang:2021yaf,Wu:2021dmf,Estienne:2021hbx}.
 \par
The universal physics of cusps can be formulated within general conformal field theory (CFT) in the presence of a conformal line defect \cite{Gaiotto:2013nva,Billo:2016cpy}. In this setting, the absence of intrinsic dynamical scales and the additional structure implied by conformal invariance allow the relevant physics to be organized particularly cleanly. Cusps form part of the local data of a line defect and are therefore constrained by several manifestations of defect locality, including the operator-state correspondence, a corresponding analog of the operator algebra of local operators \cite{Alday:2010ku,Gaiotto:2010fk,Cavaglia:2018lxi,exciting_magic}, and defect fusion rules \cite{Bachas:2007td,Bachas:2013ora,Diatlyk:2024zkk,Diatlyk:2024qpr,Kravchuk:2024qoh,Cuomo:2024psk}.
\par
Different angular limits of the cusp isolate different pieces of this structure. Near the smooth limit $\theta \to \pi$, the cusp anomalous dimension $\Gamma_{ab}(\theta)$ is governed by local defect data, including defect-changing operators (DCOs) and, for deformations of a fixed defect, the displacement operator \cite{Correa:2012at,Cuomo:2024psk}. In the fusion limit $\theta \to 0$, $\Gamma_{ab}(\theta)$ probes the spectrum and universal Wilson coefficients entering the fusion of the two line defects \cite{Bachas:2007td,Bachas:2013ora,Diatlyk:2024zkk,Diatlyk:2024qpr,Kravchuk:2024qoh,Cuomo:2024psk}. After Lorentzian continuation, the lightlike limit instead enters a generalized Sudakov regime naturally described by heavy-particle effective field theory \cite{Korchemsky:1991zp,Cuomo:2026mop}. Moreover, in gauge theories the null cusp anomalous dimension controls the asymptotic logarithmic growth of anomalous dimensions of large-spin local operators \cite{Korchemsky:1988si,Gubser:2002tv,Kruczenski:2002fb,Alday:2007mf}. Cusps thus offer a window into local and non-local structures in CFT.
\par
We therefore seek quantitative constraints on cusps in general, with the hope of gaining insight into interplay between local and non-local CFT observables. A common way to approach such goals is to study the consequences of unitarity and locality for the consistency of observables, which has a long history, more broadly speaking, in the context of conformal boundaries and defects \cite{Cardy:1986gw,Cardy:1991tv,Lewellen:1991tb,McAvity:1995zd,Liendo:2012hy,Gaiotto:2013nva,Gliozzi:2015qsa,Billo:2016cpy}. For cusps, nontrivial constraints on $\Gamma_{a\bar a}(\theta)$ follow from reflection positivity \cite{Pobylitsa:2007ky,Cuomo:2024psk,Chandra:2026wnf}, while closely related constraints on corner contributions to entanglement entropy follow from strong subadditivity \cite{Hirata:2006jx,Casini:2008as}. Other positivity constraints on Wilson loops and entanglement entropy have were explored in \cite{Hirata:2008ms,Casini:2012rn}.

These known constraints compare wedges at nearby opening angles, through applying reflection positivity through the cusp \footnote{Subtleties associated with this choice of quantization plane, together with a careful derivation of these known inequalities using the cusp operator product expansion, are discussed in \cite{exciting_magic}.}, and imply ordinary concavity-type relations for $\Gamma_{a\bar a}(\theta)$ in $\theta$. While they can technically relate widely separated angles, they are too weak to simultaneously retain the leading data of the two asymptotic regimes, such as the fusion singularity at $\theta\to0$ and the DCO data controlling the smooth limit $\theta\to\pi$. We instead accompany changes of the opening angle by conformal transformations chosen so that the resulting configurations define states in a common defect Hilbert space, allowing us to directly exploit unitarity in a complementary quantization.

% These constraints arise from comparing wedges at nearby opening angles and {\color{orange} struggle to} connect the distinct asymptotic regimes of the cusp\RL{I think the more precise statement is that they don't constrain the expansion parameters? In some sense midpoint concavity does connect different regimes, just it is very weak? Like at $\theta = \pi/2$ midpoint gives me a bound which can involve cusp angles arbitrarily close to $0,\pi$, but it's just a very weak bound} \YW{yes it cannot see the divergence in fusion limit and the DCO data at the same time. the dressing in the affine CAD weakens the singularity to make this happen}. As we show below, conformal transformations \RL{Should we be careful about calling it a "conformal transformation"? Conformal transformations preserve angles? Although yea it's true the tuning of cusp angle is given by conf. trans., maybe I'm just confused...}\YW{conformal transformations of the family of straight cusps or looking at them in different conformal frames} generate a different class of deformations, leading to a conformal notion of concavity and, in turn, stronger relations among cusp data.
\par 
In this work, we derive a new family of inequalities for the cusp anomalous dimension $\Gamma_{ab}(\theta)$ of two oriented outgoing line defects $a$ and $b$, starting from the following mother inequality, which we refer to as the \emph{conformal concavity} of $\Gamma_{ab}(\theta)$
\begin{gather}
    \cE_{ab}''(u)\le 0 \,,\label{eq:concavity}\\
    \cE_{ab}(u)\equiv \sqrt{u}\,\Gamma_{ab}(\theta(u))\,, \quad u=\tan^2{\theta\over 4}\label{aCAD}\,,
\end{gather}
where $u \in [0,1]$ parameterizes an affine family of Hamiltonians whose geometric origin will be explained shortly, and $\cE_{ab}(u)$, which we refer to as the \emph{affine cusp anomalous dimension}, fixes the ground state energy of these Hamiltonians. \par 
Together with the limiting behavior
\ie
\cE_{ab}(0)=C_{ab\bar c}/4\,,~\cE_{ab}(1)=\Delta_{ab}\,,~
\cE'_{ab}(1)=\frac{\Delta_{ab}}{2}-A_{ab}\,,
\label{eq:Esmoothconds}
\fe
some additional useful inequalities \eqref{gamma2d}, \eqref{gamma1d}, and their integrated consequences, then follow directly from \eqref{eq:concavity}. In \eqref{eq:Esmoothconds}, $\Delta_{ab}$ is the scaling dimension of the lowest defect-changing operator (DCO) $\cO_{ab}$ interpolating between $a$ and $b$, while
\ie
A_{ab}\equiv -{\Gamma_{ab}'(\theta)}|_{\theta=\pi}
\label{jf}
\fe
is the \textit{junction torque} conjugate to the bending angle $\D=\pi-\theta$.
In particular, $A_{ab}=0$ when $\cO_{ab}$ is a scalar under transverse rotations but it is generically nonzero for interfaces between $2d$ CFTs. Importantly, the affine cusp anomalous dimension $\cE_{ab}$ remains finite in the fusion limit $u\to0$, where it is determined by the dominant fusion channel $c$ through the Casimir energy coefficient $C_{ab\bar c}$ \cite{Diatlyk:2024zkk,Diatlyk:2024qpr,Kravchuk:2024qoh,Cuomo:2024psk}. As we show, conformal concavity propagates this fusion data to intermediate angles, directly linking the fusion and smooth regimes.

\begin{figure*}[t]
\centering
\includegraphics[width=0.85\linewidth]{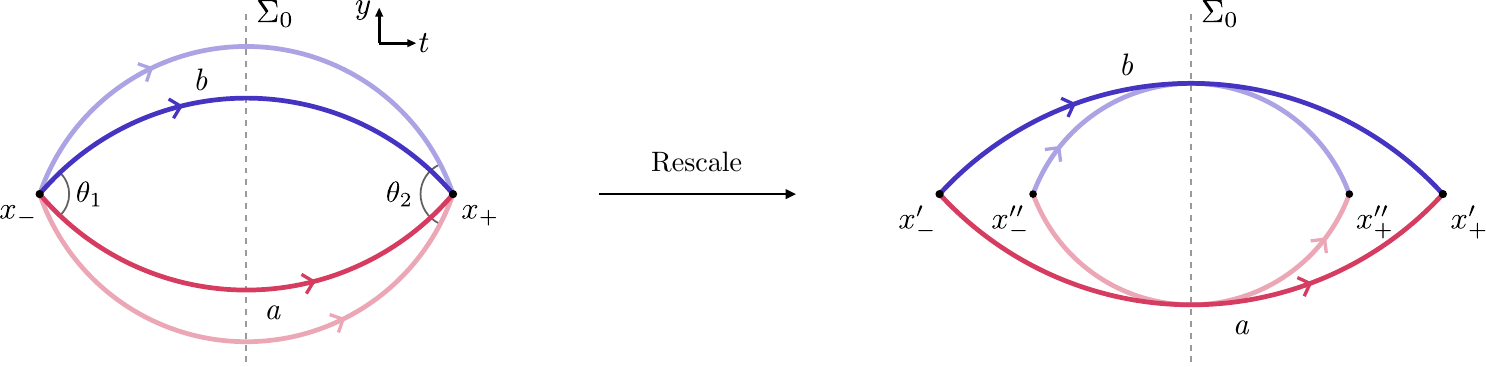}
\caption{
Conformal eye geometry in the NS frame with fixed points $x_\pm$. On the left, eyes with opening angles $\theta_1$ and $\theta_2$ define states $|\psi_{ab}(\theta_i)\rrangle$ in Hilbert spaces $\cH_{ab}(u_i)$, where the defect $a,b$ punctures on $\Sigma_0$ lie at $y=\pm\sqrt{u_i}$. We then apply a dilatation to each eye configuration so that the punctures land at $y=\pm1$, and the cusp locations land at $x_\pm'$ and $x_\pm''$, which have distances $1/\sqrt{u_1}$ and $1/\sqrt{u_2}$ from the origin, respectively, producing the states $|\psi_{ab}(\theta(u_i))\rrangle_{u_i}$ in a common defect Hilbert space $\cH_{ab}$.
}
\label{fig:eye}
\end{figure*}
The mother inequality has a simple geometric origin. Consider two cusps forming an ``eye'' whose circular ``eyelids'', illustrated in Figure~\ref{fig:eye}, follow conformal Killing flows associated with north-south (NS) quantization. The fixed points $x_\pm$ of these flows, defining future and past temporal infinity, lie at the two cusps of the eye. The key observation is that varying the cusp angle by changing the height of the circular eyelids can be compensated by a corresponding change of the NS foliation and hence of the conformal Hamiltonian on a fixed defect Hilbert space. Reflection positivity across the midplane, together with the variational principle in quantum mechanics, then directly yields \eqref{eq:concavity}.
\par
Conformal concavity is strictly stronger than the usual angular concavity of $\Gamma_{a\bar a}(\theta)$, with correspondingly broader consequences. In the fusion regime, the eye inequalities yield bounds on defect-fusion data that encompass and generically strengthen those inferred from the lightlike limit \cite{Cuomo:2026mop}. When fusion is dominated by the trivial defect, conformal concavity further constrains subleading fusion EFT coefficients and, through them, the scaling dimensions of operators exchanged in the fusion channel. Most strikingly, \eqref{eq:concavity} directly connects the fusion and smooth limits: local data controlling the smooth cusp constrain the singular fusion regime, and vice versa. In particular, we obtain bounds on the lowest defect excitation energy in terms of the Casimir coefficient, including when the lowest excitation carries nonzero transverse spin. The eye geometry thus turns the angular dependence of the cusp into a quantitative bridge between otherwise disparate sectors of defect CFT data.

\emph{Conformal geometry of the eye}---The concavity inequality \eqref{eq:concavity} follows from reflection positivity applied to a simple conformal geometry: two cusps joined by circular arcs to form an ``eye''. The cusps are formed by two oriented conformal defects $a$ and $b$ joining at an angle $\theta$. The natural quantization scheme for this defect geometry is NS quantization \cite{Rychkov:2016iqz}, which we now describe. In Appendix~\ref{app:eyeframes}, we discuss the conformally equivalent radial and cylinder frames.
\par 
Let $x=(t,x^i) \in \mathbb R^d$, with $t=x^0$ the Euclidean time, and introduce two distinguished points
\ie
x_\pm=(\pm1,0,\ldots,0)\,.
\fe
The NS conformal Killing vector generates a conformal flow with $x_\pm$ as fixed points. 
In any $2d$ plane containing the $t$ axis, the integral curves of this flow are circular arcs passing through $x_\pm$. The conformal eye is obtained by choosing two such trajectories, which form two circular ``eyelids'' intersecting at $x_+$ and $x_-$, each with opening angle $\theta$ (see Figure~\ref{fig:eye}). \par
Let us now study the quantization of this defect configuration. Following the conventions of \cite{Simmons-Duffin:2016gjk}, we define the conformal charges
\ie
Q_\xi\equiv-\int_{\Sigma_0} dS_\mu\,\xi_\nu T^{\mu\nu}\,,
\label{charge}
\fe
associated with the conformal Killing vectors
\ie
\xi_{P_0}=\partial_t\,,~
\xi_{K_0} =2t x^i\partial_i-(x^i x_i-t^2)\partial_t\,,~
\xi_D =x^\mu\partial_\mu\,.
\label{eq:CKVs} 
\fe
To ease notation, we suppress the dependence of the conformal generators on the types and positions of the defects. We define the following 
\ie
    H\equiv \frac{1}{2}(P_0-K_0),\quad
    P\equiv \frac{1}{2}( P_0+ K_0),\quad D,
\fe
which form an $\mathfrak{sl}(2, \mathbb R)$ algebra
\ie
[ D,  H]=  P\,,\quad
[ D, P]= H\,,\quad
[ H, P]=-D\,.
\fe
With respect to reflection across the NS quantization slice at $t=0$, which is a hyperplane $ \Sigma_0 \simeq \mathbb R^{d-1}$, their adjoint properties are $H^\dagger = H$, $P^\dagger = P$, and $D^\dagger = -D$ \footnote{This is to be contrasted with the adjoint properties in the more family radial quantization (see Appendix~\ref{app:eyeframes} for details).}.

As we will see, $H$ generates the conformal flow that preserves the cusp geometry, while the broken generators $D$ and $P$ play essential roles in relating different cusp configurations and the associated Hamiltonians. More broadly, broken conformal generators have recently been exploited to derive nontrivial constraints on defect observables \cite{Gabai:2025zcs,Girault:2025kzt,Kong:2025sbk,Belton:2025ief,Drukker:2025dfm,Belton:2026xaw}.

\par 
Taking the eye to lie in the $(t,y)$ plane with $y=x^1$, the two eyelids intersect the quantization slice $\Sigma_0$ orthogonally at
\ie
y=\pm y_*\,,\quad
y_*=\tan{\theta\over4}\,.
\label{ystar}
\fe
The eye geometry is then entirely generated by the flow of $H$. In particular, the state $|\psi_{ab}(\theta)\rrangle\in\cH_{ab}(y_)$ preparing the cusped defect configuration in the $t<0$ region is the ground state of $H$,
\begin{align}
H|\psi_{ab}(\theta)\rrangle
&=\Gamma_{ab}(\theta)|\psi_{ab}(\theta)\rrangle\,,
\label{eq:eyegd}
\end{align}
where $\cH_{ab}(y_*)$ denotes the Hilbert space on $\Sigma_0$ punctured by the defects $a,b$ at $y=\pm y_*$. As we now show, $P$ and $D$ generate nontrivial deformations of the eye, corresponding respectively to a squeeze in the $t$ direction and an overall dilation. These deformations will be crucial below.
\par
\emph{Conformal concavity from variational principle}---
On the quantization slice $\Sigma_0$, the conformal Killing vectors generated by $P_0$ and $K_0$ have no tangential components, and therefore do not move the defect punctures within $\Sigma_0$. The same is true for any linear combination of them, in particular $H$ and $P$. This allows the corresponding family of Hamiltonians to be compared on a common defect Hilbert space. We use a conformal transformation (a dilation for the symmetric eye in Figure~\ref{fig:eye}) to place the two defect punctures at $y=\pm1$, and denote this fixed Hilbert space by $\cH_{ab}$.

It is then natural to consider the one-parameter affine family of Hamiltonians on $\cH_{ab}$,
\ie
H(u)\equiv {P_0-uK_0\over 2}
={ (1+u)H+(1-u)P\over 2}\,.
\label{Hu}
\fe
The geometric meaning of this family follows directly from the eye construction: its ground state is a cusp whose opening angle varies continuously with $u$, with the cusp located at a fixed point of the conformal Killing vector field generated by $H(u)$. To see this, let us introduce
\ie
u=\tan^2{\theta\over4}\,,\quad s=-{1\over2}\log u\,,
\fe
so that the corresponding half-eye of opening angle $\theta=\theta(u)$ intersects $\Sigma_0$ at
\ie
y=\pm y_*=\pm\sqrt{u}\,.
\fe
The dilation $e^{sD}$ sends these punctures to $y=\pm1$ and therefore maps the half-eye state into the fixed Hilbert space $\cH_{ab}$. Since $D^\dagger=-D$, this transformation is unitary. Using $[D,P_0]=P_0$ and $[D,K_0]=-K_0$, we find
\ie
H(u)=\sqrt{u}e^{sD}He^{-sD}\,.
\label{Hconjugate}
\fe
It follows that the ground state of $H(u)$ is precisely the dilated half-eye state,
\ie
|\psi_{ab}(\theta(u))\rrangle_u
=e^{sD}|\psi_{ab}(\theta(u))\rrangle\,,
\label{Omegau}
\fe
and combining \eqref{Hconjugate} with \eqref{eq:eyegd} gives
\ie
H(u)|\psi_{ab}(\theta(u))\rrangle_u
=\cE_{ab}(u)|\psi_{ab}(\theta(u))\rrangle_u\,.
\label{eq:Eground}
\fe
Thus, the affine cusp anomalous dimension $\cE_{ab}(u)$ is precisely the ground-state energy of an affine family of Hamiltonians acting on the fixed defect Hilbert space $\cH_{ab}$. In particular, the factor $\sqrt{u}$ in \eqref{aCAD} arises from the rescaling of the Hamiltonian under the dilation that brings the punctures to fixed positions. See Figure~\ref{fig:eye} for an illustration.

Conformal concavity now follows immediately from the variational principle. Namely,
\ie
\cE_{ab}(u)=\inf_{|\psi\ra\in\cH_{ab}}
{\la\psi|H(u)|\psi\ra\over\la\psi|\psi\ra}\,.
\label{Rayleigh}
\fe
Since $H(u)$ depends affinely on $u$, for $u_1,u_2\in[0,1]$ and $\lambda\in[0,1]$,
\ie
\lambda\,\cE_{ab}(u_1)+(1-\lambda)\cE_{ab}(u_2)
\leq
\cE_{ab}\big(\lambda u_1+(1-\lambda)u_2\big)\,.
\label{eq:concavefinite}
\fe
This finite concavity inequality is the general statement. Whenever $\cE_{ab}(u)$ is twice differentiable, it is equivalent to \eqref{eq:concavity}. Physically, the marginal energetic cost of opening the eye decreases as the opening grows, when measured in the natural affine variable $u$. 

Here we have focused on the planar, reflection-symmetric eye geometry. One may ask whether more general conformal deformations, including non-planar eyes, lead to additional concavity properties of the cusp. As we explain in Appendix~\ref{app:nonplane}, the resulting inequalities are consistent but weaker than the mother inequality \eqref{eq:concavity}.

\par
\emph{Daughter inequalities and sum rule}---Let us now discuss some immediate consequences of the mother inequality \eqref{eq:concavity}, or alternatively its finite form \eqref{eq:concavefinite}, and compare them with the ordinary angular concavity constraints.

Expressed in terms of the opening angle $\theta$, conformal concavity amounts to
\ie
\Gamma_{ab}(\theta)-\sin\theta\,\Gamma'_{ab}(\theta)
-4\sin^2{\theta\over2}\Gamma''_{ab}(\theta)\geq0\,.
\label{gamma2d}
\fe
A useful daughter inequality follows from the tangent line inequality for a concave function,
\ie
\cE_{ab}(1)\leq\cE_{ab}(u)+(1-u)\cE'_{ab}(u)\,.
\fe
Using $\cE_{ab}(1)=\Delta_{ab}$ and translating back to the angular variable $\theta$, we obtain
\ie
\Gamma_{ab}(\theta)+\sin\theta\,\Gamma'_{ab}(\theta)
\geq\Delta_{ab}\sin{\theta\over2}\,.
\label{gamma1d}
\fe

For $b=\bar a$, the cusp is obtained by bending a single line defect and the smooth junction is the identity, so that $\Delta_{a\bar a}=0$. The eye inequalities \eqref{gamma1d} and \eqref{gamma2d} then provide strictly stronger constraints than the familiar consequences of ordinary angular concavity \cite{Cuomo:2024psk},
\ie
\Gamma''_{a\bar a}(\theta)\leq0\,.
\fe
In the $\theta \to \pi$ limit we have $\Gamma_{a \bar a}(\pi) = \Gamma'_{a \bar a}(\pi) = 0$, so ordinary concavity implies
\ie
\Gamma_{a\bar a}(\theta)\leq0\,,\quad
\Gamma'_{a\bar a}(\theta)\geq0\,.
\fe
These constraints follow from reflection positivity of connected two-point functions of the displacement operator integrated along the two rays of a single cusp \cite{Cuomo:2024psk}. 
\par
The conformal concavity admits a sharper spectral form. Assuming for simplicity a nondegenerate ground state in $\cH_{ab}$, standard second-order perturbation theory with perturbing ``potential'' $-K_0/2$ 
gives
\ie
\cE_{ab}''(u)
=-{1\over2}\sum_{n\geq1}
{|\la n|K_0|0\ra|^2\over E_n-E_0}
\leq0\,.
\label{eyesumrule}
\fe
Here $|0\ra=|\psi_{ab}(\theta(u))\ra_u$ is the ground state of $H(u)$ with energy $E_0=\cE_{ab}(u)$, while $|n\ra$ are orthonormal excited states with energies $E_n>E_0$. Thus, $\cE_{ab}''(u)$ directly measures the spectral weight with which $K_0$ connects the cusp ground state to excited cusp states.

In particular, saturation at a given $u$ requires
\ie
\la n|K_0|0\ra=0\,,\quad n\geq1\,. 
\fe
Furthermore, the asymptotic values \eqref{eq:Esmoothconds} fixes in this case
\ie 
\cE_{ab}(u)={C_{ab\bar c} (1-u) \over 4} +\Delta_{ab}\,,\quad A_{ab}={C_{ab\bar c}-2 \Delta_{ab} \over 4}\,.
\label{saturation}
\fe
In particular, for $b=\bar a$, this requires $\cE_{a\bar a}(u)=0$ and thus the line defect is topological. Therefore, for cusp on a single line defect, strict conformal concavity therefore quantifies the departure from a topological cusp through nontrivial matrix elements of $K_0$ between the cusp ground state and excited cusp states.

\par 
\emph{Fusion and Bremsstrahlung limits}---
The best-understood regimes of $\Gamma_{ab}(\theta)$ are the limits $\theta\to0$ and $\theta\to\pi$, corresponding respectively to the fusion and generalized Bremsstrahlung (or smooth) limits. In the fusion limit, a nearby pair of defects can often be described by an EFT built from irrelevant perturbations of another line-defect fixed point of the theory \cite{Bachas:2007td,Bachas:2013ora,Kravchuk:2024qoh}, although other possibilities can arise \cite{Chernikov:2026lcv}. In the generalized Bremsstrahlung limit, the small deflection from a smooth line can instead be treated perturbatively as a deformation generated by the displacement operator $\mathbb D_i(x(\tau))$. We briefly review the physics of these two limits and the useful quantities they define, which will subsequently be constrained by conformal concavity.
\par 
We start with the fusion limit, which has been explained in detail in \cite{Kravchuk:2024qoh,Cuomo:2024psk} whose result for $\Gamma_{ab}(\theta)$ we quote: as $\theta \to 0$ we expect
\ie 
\Gamma_{ab}(\theta)={C_{a b \bar c}\over \theta} +\Delta_{c\id} +a_2 \theta + \A \theta^p+ \dots\,,
\label{fusionlimit}
\fe
where $c$ is the leading fusion channel of $a\times b$, meaning $C_{ab\bar c}$ is the smallest among the possible fusion channels, $\Delta_{c\id}$ is the dimension of the leading symmetry-allowed endpoint primary operator of $c$, $a_2$ is the leading curvature correction coefficient from the fusion EFT, and the last term takes into account the correction from the leading irrelevant operator of dimension $\Delta_{\rm irr}$ on the fused defect $c$ with $p=2(\Delta_{\rm irr}-1)$ if $c=\id$ is the identity defect (and thus $\Delta_{c\id}=0$) and more generally $p=\Delta_{\rm irr}-1$. 
\par
The Bremsstrahlung limit $\theta\to\pi$ has been studied extensively for conjugate line defects, $a=\bar b$. Here we consider the more general setting in which the smooth configuration supports a nontrivial DCO $\cO_{ab}$. In this limit, the response to a small deflection is controlled by correlation functions of the displacement operator in the state $|\cO_{ab}\ra$ created by the DCO on the smooth line defect:
\ie
\Gamma_{ab}(\pi-\delta)=\Delta_{ab}+A_{ab}\D-B_{ab} \D^2+\dots\,.
\label{smoothlimit}
\fe
Here $A_{ab}$ and $B_{ab}$ characterize, respectively, the linear and quadratic responses to a small deformation of the junction. In particular,
\ie
A_{ab}={\la \cO_{ab} | \mathbb D_\perp |\cO_{ab}\ra\over \la \cO_{ab} |\cO_{ab}\ra}=-{\Gamma_{ab}'(\theta)}|_{\theta=\pi}\,,
\fe
where $\mathbb D_\perp=\hat n\cdot\mathbb D$ is the component of the displacement operator that generates the cusp deformation of the smooth defect contour in the transverse direction $\hat n$.

When the smooth line defect preserves the transverse reflection $\hat n\to-\hat n$, the two directions of deformation are equivalent. Since $\Gamma_{ab}$ is the ground-state energy, its one-sided linear response away from the smooth configuration must satisfy $A_{ab}\leq0$. A nonzero linear response in this case requires a degeneracy of DCO states at the smooth configuration, which can be split at linear order by the small angle deformation $\D$. If, in addition, the DCO is itself invariant under the reflection (e.g. if it is a scalar under transverse rotations), then the one-point function of $\mathbb D_\perp$ vanishes and hence $A_{ab}=0$.

The linear term in \eqref{smoothlimit} can nevertheless be nonzero, even for a trivial DCO, when the transverse reflection symmetry is absent for the line defect. For example, for factorized interfaces in $2d$ CFT, $A_{ab}$ receives a contribution from the mismatch between the CFT ground-state (Casimir) energies on the two sides and can therefore have either sign when $c_1\neq c_2$. More generally, $A_{ab}$ can be nonzero for line defects embedded in higher-dimensional conformal defects that break the relevant transverse reflection symmetry.

When $a=\bar b$ and $A_{ab}=0$, $B_{a\bar a}$ is the Bremsstrahlung coefficient which measures the braking radiation of the probe particle representing the defect $a$ under sudden deflection. In CFT, this coefficient is determined by the displacement two-point function on the defect $a$,
\ie 
B_{a\bar a} ={C^a_\mathbb{D}\over 12}\,,~~\la 0_a|  \mathbb D_i(t) \mathbb D_j(0) |0_a\ra ={C^a_\mathbb{D}\D_{ij}\over t^4}\,,
\fe
and thus obviously positive by reflection positivity, where we let $|0_a\rangle$ represent the Minkowski vacuum in the presence of the defect $a$. It is known exactly from localization in $\mathcal{N}\geq 2$ supersymmetric theories \cite{Correa:2012at,Lewkowycz:2013laa,Fiol:2015spa,Bianchi:2018zpb}. 

In the more general Lorentzian setting, the coefficients in \eqref{smoothlimit} characterize an impurity undergoing a simultaneous quench and sudden kick: both the type of the probe and its velocity change at the same instant. In particular, the nonzero $\Delta_{ab}$ is the orthogonality exponent associated with the quench $a\to b$, familiar from the Anderson orthogonality catastrophe \cite{Affleck_1994,Affleck:1996mm}. The coefficient $A_{ab}$ is a junction torque governing the linear response to the kick, while $B_{ab}$ governs the quadratic recoil response and generalizes the usual Bremsstrahlung coefficient recovered for $b=\bar a$ \cite{Billo:2016cpy}.

\begin{table*}[t]
\caption{\label{table:IsingConcavity}Fuzzy-sphere test of the conformal concavity for the 3d Ising pinning defect cusp \cite{Cuomo:2024psk}. Here $s_i$ is the secant slope and they are monotonically decreasing as expected for the conformal concavity. We have omitted the error bars in this Table.\footnote{In \cite{Cuomo:2024psk},  error bars for the cusp anomalous dimensions were estimated based on finite-size and pinning-strength variation. However they do not provide the full inter-angle covariance matrix. Since each slope $s_i$ depends on two neighboring cusp values and adjacent slopes share data,  these correlations will be important for a proper uncertainty or significance analysis for the ordered $s_i$ presented in the Table.}}
\begin{ruledtabular}
\begin{tabular}{c c c @{\hspace{10mm}} c c c}
\multicolumn{3}{c}{aligned $(++)$} & \multicolumn{3}{c}{anti-aligned $(+-)$}\\\hline
$\theta_i/\pi$ & $\Gamma_{++}$ & $s_i^{++}$ &
$\theta_i/\pi$ & $\Gamma_{+-}$ & $s_i^{+-}$\\ \hline
.3 & $-.1900$ & $.1924$ & .4 & $1.2900$ & $.5654$\\
.4 & $-.1120$ & $.1309$ & .5 & $1.1020$ & $.5332$\\
.5 & $-.0670$ & $.0895$ & .6 & $.9880$  & $.4997$\\
.6 & $-.0390$ & $.0657$ & .7 & $.9160$  & $.4741$\\
.7 & $-.0200$ & $.0399$ & .8 & $.8720$  & $.4487$\\
.8 & $-.0085$ & $.0217$ & .9 & $.8477$  & ---\\
.9 & $-.0021$ & ---     &    &          &
\end{tabular}
\end{ruledtabular}
\end{table*}
\emph{Bounds on fusion EFT}---The conformal concavity inequalities impose nontrivial constraints on the spectrum and couplings of DCOs beyond the standard unitarity bounds. Combining the smooth limit and the fusion limit values \eqref{eq:Esmoothconds}, conformal concavity implies
\ie
\Delta_{ab}\geq \frac{C_{ab\bar c}}{2}-2A_{ab}\,.
\label{eq:boundonDCO}
\fe
Equality can occur only when the concavity bound is saturated throughout the entire interval as in \eqref{saturation}.
\par
The bound \eqref{eq:boundonDCO} can be substantially stronger than the ordinary unitarity bound $\Delta_{ab} \ge 0$. In many nontrivial cases $C_{ab\bar c}\geq0$, while transverse reflection symmetry implies $A_{ab}\leq0$, so both terms on the right-hand side strengthen the lower bound. Further, the bound gives a striking relation between the defect fusion data and the stability of non-simple defects: supposing $a$ and $b$ are individually stable, $a \times b$ being sufficiently heavy (i.e. large enough $C_{ab\bar c}>0$) leads to the perturbative stability of $a \oplus b$! \par
As a concrete test of \eqref{eq:boundonDCO} in an interacting example, consider a cusp formed by opposite pinning-field defects in the $3d$ Ising CFT. Using the fuzzy-sphere result $C_{+-\id} = 1.4(2)$ \cite{Cuomo:2024psk}, combined with $A_{+-}= 0$, the bound gives
\ie
\Delta_{+-}\geq \frac{C_{+-\id}}{2}=0.70(10)\,,
\fe
which is comfortably satisfied by the measured values from fuzzy -phere and conformal bootstrap \cite{Zhou:2023fqu,Lanzetta:2025xfw}. \par 
A nice illustration of the power of the conformal concavity arises when the fusion limit
$a\times b$ is dominated by the identity defect $\id$, so that
$\Delta_{c\id}=0$.  We write
\ie
\Gamma_{ab}(\theta)
=
\frac{C_{ab\id}}{\theta}
+a_2\theta
+\A\,\theta^{2(\Delta_{\rm irr}-1)}
+a_4\theta^3+\dots\,,
\label{fusionidentityexp}
\fe
where $\Delta_{\rm irr}$ in this case is simply the dimension of the lightest bulk operator that participate in the fusion process (irrelevant on the trivial defect $\id$) and $a_2,a_4$ are the leading and subleading curvature corrections from the nontrivial cusp geometry \cite{Kravchuk:2024qoh}. In contrast to the ordinary angular concavity $\Gamma_{a\bar a}''(\theta)\leq 0$, which near $\theta \to 0$ is dominated by the Casimir term and so only fixes the its sign $C_{a\bar a\bar c}\geq 0$, the conformal concavity condition is more powerful because the differential operator annihilates the leading singularity in \eqref{fusionidentityexp}. \par
\begin{figure}[t]
  \centering
  \includegraphics{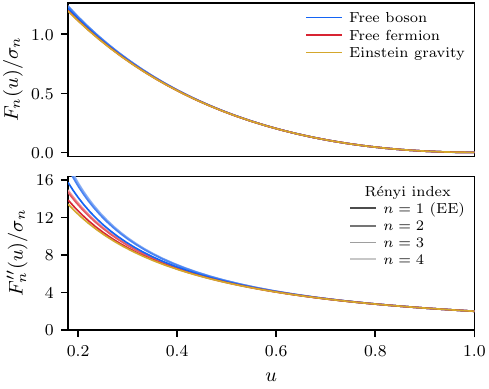}
  \caption{Affine corner contributions to Rényi entropies $F_n(u)$, defined in \eqref{dressedcorner}, as well as their second derivatives $F''_n(u)$ normalized by $\sigma_n$ for a free real scalar \cite{Casini:2006hu}, free Dirac fermion \cite{Casini:2008as}, and Einstein gravity \cite{Bueno:2015xda}. For Einstein gravity we only have access to $n=1$, i.e. the entanglement entropy. }
  \label{fig:entropycorner}
\end{figure}
Indeed, conformal concavity \eqref{gamma2d} applied to the fusion limit implies that for $1<\Delta_{\rm irr}<5/2$ \footnote{The upper restriction is so that the analytic $a_4\theta^4$ term from a four-derivative curvature correction in the fusion EFT \cite{Kravchuk:2024qoh} are subleading.}, the following interesting correlation between the sign of the irrelevant correction to the cusp and a threshold dimension at $3/2$ for bulk operators that corrects the fusion EFT
\ie 
\A(\Delta_{\rm irr}-3/2)\leq 0\,.
\fe
In the interacting ${\rm O}(N)$ CFT, for the anti-aligned pinning cusp, the leading irrelevant operator that corrects the fusion limit is from the leading rank two symmetric tensor operator $\cT$ in the bulk CFT, which is just the energy operator in the Ising case. From conformal bootstrap and large $N$ analysis, this operator has dimension $\Delta=1.41262528(29)$ \cite{Chang:2024whx} and decreases towards $\Delta_{N\gg 1}=1$ with a positive ${1\over N}$ correction. Therefore we conclude immediately that $\A_{+-}\geq 0$
which can be potentially tested by more refined fuzzy-sphere analysis and verified in the large $N$ limit.  \par 
For $\Delta_{\rm irr}>3/2$, the eye inequality \eqref{gamma1d} further produces the following constraint involving the Wilson coefficient $a_2$ of the leading curvature term in the fusion EFT,
\ie 
C_{ab\id}
+12 a_2
-3\Delta_{ab}
\geq0\,,
\label{CaD}
\fe 
which also gives an upper bound on the lowest DCO dimension in terms of the fusion data (compared to \eqref{eq:boundonDCO}).

For $b=\bar a$, \eqref{CaD} becomes a simple direct constraint between the first two Wilson coefficients $a_0\equiv -C_{a\bar a \bar c}$ and $a_2$ accessible in the cusp geometry,
\ie 
\Delta_{\rm irr}>{3\over 2}\Rightarrow a_2\geq {a_0\over 12}\,.
\label{a0a2bound}
\fe
The condition holds for the BPS Wilson loops in the $\cN=4$ SYM, and one can easily confirm this inequality explicitly in this case using \eqref{planarSYM} in the End Matter. This inequality must also hold for any conformal line defects in any interacting $5d$ CFT due to the stronger unitarity bound on $\Delta$ \cite{Minwalla:1997ka}. \par 
\emph{Bounds and sum rule for generalized Bremsstrahlung}---To better understand the generalized Bremsstrahlung process, it is useful to regard the line defects as heavy particles coupled to a gapless bath described by the CFT \cite{Cuomo:2026mop}. For ordinary Bremsstrahlung, the leading quadratic coefficient $B_{a\bar a}$ in the smooth limit measures the soft radiation from accelerating a fixed impurity $a$, and is positive by unitarity. When the kick is accompanied by a quench $a\to b$, however, the soft radiation induced by recoil can interfere constructively or destructively with that generated by the quench. Consequently, $B_{ab}$ need not be positive: a negative value means that a small recoil weakens, rather than strengthens, the infrared singularity already present in the static quench. An explicit example is provided by factorized interfaces in 2d CFT \cite{Cardy:1988tk}, as discussed in the End Matter. Explicitly, this loss of positivity is reflected in the mixed two-point functions of displacement operators inserted on the two defect legs.

The conformal concavity \eqref{eq:concavity} applied at the smooth end $u=1$  
instead bounds the amount of destructive interference that can occur. In the smooth limit, it yields the positive combination
\ie
\Delta_{ab} + 8B_{ab}=-4\cE''_{ab}(1)=2\sum_{n\geq 1}{|\la n | K_0|\cO_{ab}\ra|^2\over E_n-\Delta_{ab}}\geq 0\,.
\fe 
This inequality generalizes the usual Bremsstrahlung positivity to the case of a defect-changing quench. \par 
\emph{Further tests of conformal concavity}---Here we check our eye inequalities by testing them against known examples of $\Gamma_{ab}(\theta)$ at general angles. For examples where there are explicit formulas for $\Gamma_{ab}(\theta)$, it is straightforward to verify that the inequalities are satisfied. Such examples include factorized interfaces in $2d$ CFT, the pinning field in $4d$ free scalar theory, the Wilson line in $4d$ Maxwell theory, and the $1/2$-BPS fundamental Wilson line in planar $\mathcal N=4$ $SU(N)$ SYM. We provide the formulas and assess their conformal concavity in the End Matter. We focus here on some more interesting tests, where the full $\theta$-dependence of $\Gamma_{ab}(\theta)$ is not exactly computable. 

We first consider the pinning field of the $3d$ Ising CFT. As mentioned, the cusp anomalous dimensions $\Gamma_{++}(\theta)$ and $\Gamma_{+-}(\theta)$ of the pinning defects with positive and negative couplings to the spin operator in the $3d$ Ising CFT have been obtained numerically from fuzzy-sphere regularization for a discrete set of angles \cite{Cuomo:2024psk}, summarized in Table~\ref{table:IsingConcavity}. \par
To test conformal concavity, we consider the discrete version of \eqref{eq:concavity}: for a collection of increasing angles $0\leq \theta_1\leq \theta_2\leq...\leq \theta_{N} \leq\pi$, defining the secant slope of the dressed cusp anomalous dimension $\cE_i\equiv \cE_{ab}(u_i)$ with $u_i\equiv u(\theta_i)$,
\ie 
s_i\equiv {\cE_{i+1}-\cE_{i}\over u_{i+1}-u_i}\,,
\label{secantslope}
\fe
then concavity implies $s_{1}\geq s_2 \geq\dots\geq s_{N-1}$. In Table~\ref{table:IsingConcavity}, we perform this test using the fuzzy -phere results for the pinning defect cusp in the Ising CFT, finding consistency with conformal concavity.

We next consider the case of corner contributions to Rényi and entanglement entropies in $3d$ CFT. The $n$-th Rényi entropy of $3d$ CFT $\cT$ is obtained from the free energy of the cyclic $\mZ_n$ monodromy 
defect in the $n$-th replicated CFT $\cT^{\otimes n}$ \cite{Bianchi:2015liz}. The defect location is identified with the boundary of the
entangling region. When the entangling region contains corners, the universal corner correction to the Rényi entropy $a_n(\theta)$ is precisely governed by the cusp anomalous dimension of the associated twist defect via 
\ie 
\Gamma^{(n)}_{\rm twist}(\theta)=(1-n)a_n(\theta)\,.
\fe
Conformal concavity implies that for integer $n\geq2$, the following affine corner Rényi entropy,
\ie 
F_n(u)\equiv \tan{\theta\over 4}a_n(\theta)\,, 
\label{dressedcorner}
\fe
is positive, monotonically decreasing and convex in $u=\tan^2{\theta\over 4}$. The usual corner entanglement entropy can be obtained from the $n\to 1^+$ limit of $a_n(\theta)$. While our eye inequalities do not a-priori apply to this limit, we can see that $F_{n \to 1}(u)$, i.e. the affine corner entanglement entropies, continue to satisfy \eqref{eq:concavity}. We note that strong subadditivity applied to entangling regions involving relatively boosted wedges does, however, give a bound that non-trivially mixes $a_{n\to 1}'(\theta)$ and $a_{n \to 1}''(\theta)$, which has a somewhat similar flavor to conformal concavity \cite{Casini:2008as}. \par

In Figure~\ref{fig:entropycorner}, we use the known exact results to plot affine corner Rényi entropies of free scalar and free Dirac fermion as well as the affine corner entanglement entropy for CFT dual to Einstein gravity \cite{Casini:2006hu,Casini:2008as,Bueno:2015xda}. Conformal concavity is clearly evident from the numerical plots. To facilitate the comparison, we have normalized the affine corner Rényi entropy by the leading coefficient $\sigma_{n}$ in the near smooth limit,
\ie 
a_{n}(\pi-\delta)=\sigma_{n} \D^2 +\dots\,, 
\fe
which for $n\to 1$ is also related to the stress tensor two-point function coefficient by $\lim_{n\to 1} \sigma_n = \pi^2C_T/24$ \cite{Bianchi:2015liz}.

\emph{Discussion}---In this Letter we derived new non-perturbative inequalities on cusp anomalous dimensions by applying reflection positivity to an ``eye" geometry. Our inequalities take the form of a simple geometric inequality, which we referred to as conformal concavity. As compared to previous concavity bounds, our conformal concavity bounds are much stronger, constraining subleading data in the fusion limits, and applying to cusps of unequal line defects.  We performed numerous tests of conformal concavity, and applied it to derive new results in the fusion and bremsstrahlung limits. \par 
There are a number of directions that would be interesting to pursue. In $\mathcal N=4$ SYM, the spectrum of excited cusps is accessible from integrability, making it possible to test and refine our conformal concavity bounds at finite coupling \cite{Gromov:2012eu,Gromov:2015dfa,Grabner:2020nis,Cavaglia:2018lxi,Chernikov:2026lcv}. It would also be valuable to understand whether there are 
stronger constraints on cusps. In this regard, the complete monotonicity properties observed for cusps in gauge theories \cite{Henn:2024qwe} and for corner entropies \cite{Casini:2012rn,Bueno:2015ofa} are particularly suggestive. Finally, the relations uncovered here between defect fusion and the spectrum of defect-changing operators may provide useful input for bootstrapping conformal gauge theories using line defects \cite{Chester:2016wrc,Li:2018lyb,Li:2020bnb,He:2021xvg,Albayrak:2021xtd,Lanzetta:2025xfw}.

\emph{Acknowledgments}---We thank Meng Cheng, Gabriel Cuomo, Petr Kravchuk, Juan Maldacena, and Amit Sever for useful discussions. I.M. is supported by the DOE Early Career Award DE-SC0025581, the Sloan Foundation, and the Simons Collaboration on Confinement and QCD Strings. The work of YW was supported in part by the BSF Grant 2024187 and the NSF Grant 2609962.  YW thanks the Institute for Advanced Study for hospitality during the completion of this work. Research at Perimeter Institute is supported in part by the Government of Canada through the Department of Innovation, Science and Industry Canada and by the Province of Ontario through the Ministry of Colleges and Universities. \par\vspace{1mm}\noindent
\textbf{Note added}: We thank Lorenzo Bianchi, Andrea Cavaglià, Stefanos Kousvos, Marco Meineri, and Abner David Gutierrez Romero for coordinating the simultaneous arXiv submission of their work \cite{marco} on related topics.
\section*{End Matter}
Here we discuss examples of $\Gamma_{ab}(\theta)$ with exactly known expressions and test their conformal concavity.
\subsubsection{Factorized interfaces in $2d$}
Consider factorized interfaces between two $2d$ CFTs of central charges $c_1$ and $c_2$ respectively in terms of their Cardy boundaries. The cusp anomalous dimension follows from the works \cite{Cardy:1986gw,Cardy:1988tk},
\ie 
\Gamma_{ab}^{\rm fact}(\theta)=&{c_1 \theta+ c_2 (2\pi -\theta)\over 24\pi}+ \pi\left({h_1-{c_1\over 24}\over \theta}+{h_2-{c_2\over 24}\over 2\pi-\theta}\right)\,,
\fe
where $a$ and $b$ denote two factorized interfaces and $h_1,h_2$ are the corresponding dimensions of the boundary changing operators on the two sides, so that
\ie 
\Delta_{ab}=h_1+h_2\,,~~A_{ab}={h_1-h_2\over \pi}+{c_2-c_1\over 12\pi}\,.
\fe
The conformal concavity \eqref{eq:concavity} can be verified by elementary inequalities for arbitrary $c_1,c_2,h_1,h_2\geq 0$ and saturated at $c_2=h_2=0$ and $\theta=0$.
\subsubsection{Scalar Wilson line in $4d$ free scalar theory} Scalar Wilson line (also known as pinning field defect) in the $4d$ free theory of $N$ scalars has an exactly marginal coupling $\vec g \in \mR^N$. Consider the cusp formed by such Wilson lines $a$ and $b$ with couplings $\vec g_1$ and $\vec g_2$, where we have
\ie 
\Gamma_{ab}^{\rm scalar}(\theta)=-{\vec g_1\cdot \vec g_2\over 4\pi^2}{\pi -\theta\over \sin\theta}+{|\vec g_1|^2+|\vec g_2|^2\over 8\pi^2}\,,
\label{scalarcusp}
\fe
and
\ie 
\Delta_{ab}={|\vec g_1-\vec g_2|^2\over 8\pi^2}\,,\quad A_{ab}=0\,.
\fe
The conformal concavity \eqref{eq:concavity} is verified easily and saturated precisely at $\theta=0$ and $\vec g_2=-\vec g_1$.
\subsubsection{Wilson line in $4d$ Maxwell theory}
Another free theory example is cusp formed by Wilson lines of charge $q\in \mZ$ in the $4d$ Maxwell theory. Note that Wilson lines of different charges do not form a cusp since the two defect Hilbert space is empty by Gauss's law 
\ie 
\Gamma_{a\bar a}^{\rm vector}(\theta)=-{e^2 q^2\over 4\pi}((\pi-\theta) \cot \theta+1)\,,
\label{vectorcusp}
\fe
with obviously $\Delta_{a\bar a}=0$. It is again easy to check that \eqref{eq:concavity} is satisfied and it is saturated only at $\theta=\pi$.
\subsubsection{Fundamental Wilson line in $4d$ $\cN=4$ SYM}
Here we consider the half-BPS fundamental Wilson loops in the $\cN=4$ $SU(N)$ SYM \cite{Rey:1998ik,Maldacena:1998im,Drukker:2000rr,Drukker:2011za}, specified by the coupling to a normalized combination of the 6 adjoint scalars in the theory (like in the scalar Wilson line example) and correspond to a polarization with respect to the $SO(6)_R$ symmetry. For two such Wilson lines, up to R-symmetry conjugation, the cusp anomalous dimension $\Gamma_{\rm SYM}(\theta,\phi)$ depends on the relative angle $\phi$ between the $SO(6)_R$ polarizations in additional to $\theta$.
In perturbation theory, the leading contribution comes from putting together the scalar result \eqref{scalarcusp} and the vector result \eqref{vectorcusp},
\ie 
\Gamma_{\rm SYM}^{\rm 1-loop}(\theta,\phi)=-{g_{\rm YM}^2 (N^2-1)\over 8 N\pi^2}\left({\pi -\theta\over \sin\theta} (\cos \theta+\cos \phi)\right),
\fe
and
\ie 
\Delta_{\rm SYM}(\phi)={(N^2-1) g_{\rm YM}^2\over 4 N\pi^2}\sin^2{\phi\over 2}\,,\quad 
\fe
The concavity \eqref{eq:concavity} is obviously satisfied and saturation happens at $\phi=\pi$ and $\theta=0$ as indicated by the free scalar result. 

More interestingly, the SYM cusp anomalous dimension is also known in the planar limit at strong 't Hooft coupling $\lambda=g_{\rm YM}^2 N$ \cite{Drukker:2011za}. For simplicity, let us consider the $\phi=0$ case, with $0\leq m\leq 1/2$ covering the whole range from $\theta=\pi$ to $\theta=0$ faithfully,
\ie 
 &\Gamma^{\rm strong}_{\rm SYM}(\theta) =\sqrt{\lambda\over 1-2m} {(1-m)K(m)-E(m)\over \pi}\,,\quad 
 \\
 &\theta=2\sqrt{m(1-2m)\over 1-m}\left(
 \Pi(1-m|m)-K(m)\right)\,,
 \label{planarSYM}
\fe 
where $K(m),E(m),\Pi(m|n)$ are complete Elliptic integrals of first, second and third kind respectively. It is easy to confirm numerically that \eqref{eq:concavity} is satisfied using \textit{Mathematica} and never saturated in this case.

\appendix

\section{Conformal Frames for the Eye}
\label{app:eyeframes}

\begin{figure*}[t]
\centering
\includegraphics{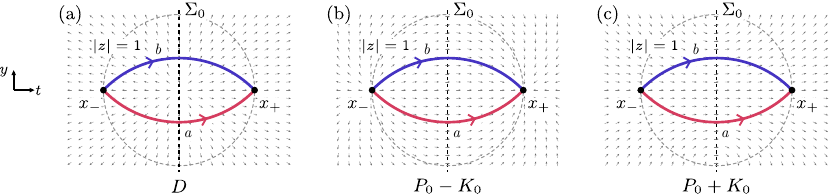}
\caption{
The vector fields corresponding to the $\mf{sl}(2,\mathbb{R})$ generators
$D$, $P_0-K_0$, and $P_0+K_0$ in the $t$--$y$ plane.
The blue and red  curves are the defect lines $b$ and $a$, respectively,
forming the eyelids of a symmetric eye in the NS frame. The vertical dotted
line denotes the NS quantization slice $\Sigma_0$, and the dashed curve is
the reference unit circle $|z|=1$.
Along $\Sigma_0$, $P_0-K_0$ and $P_0+K_0$ are normal to $\Sigma_0$,
whereas $D$ is tangent to it. On the unit circle, $D$ and $P_0+K_0$
are normal to the circle, whereas $P_0-K_0$ is tangent.
}
\label{fig:sl2vectorfields}
\end{figure*}
 
In the main text we worked in the conformal frame adapted to the NS quantization, where the affine Hamiltonian family underlying conformal concavity is most transparent after fixing the defect punctures on the quantization slice. Here we briefly relate this description to the radial and cylinder frames.

Let $z=t+iy$ denote the complex coordinate on the plane containing the eye. Before fixing the punctures, the cusp points lie at $z=\pm1$, while the two defect legs intersect $\Sigma_0$ at
\ie
z=\pm i\sqrt u\,,\quad
u=\tan^2{\theta\over4}\,.
\fe
The NS frame we choose is obtained by the dilation $z\to z/\sqrt u$, which fixes the punctures at $z=\pm i$ while moving the cusp points to $z=\pm1/\sqrt u$. The corresponding Hamiltonian \eqref{Hu} then acts on a fixed defect Hilbert space $\cH_{ab}$.

\begin{figure*}[t]
\centering
\includegraphics{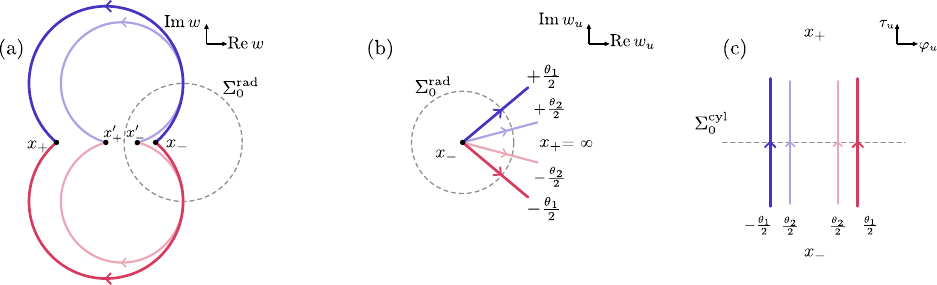}
\caption{
Conformal frames for the eye:
(a)  radial frame with fixed defect punctures but varying cusp locations
(b)  radial frame with fixed location of the cusp but varying angles 
(c)  cylinder frame.
The solid and dashed curves correspond to the opening angles $\theta_1$ and $\theta_2$ for cusps represented in different conformal frames.
}
\label{fig:eyeframes}
\end{figure*}

A $u$-independent passage to radial quantization is provided by
\ie
w={1+z\over1-z}\,.
\fe
It maps $\Sigma_0$ to the unit circle $|w|=1$, keeps the punctures fixed at $w=\pm i$, and sends the cusp points to
\ie
x_-(u)=-r(u)\,,\quad
x_+(u)=-{1\over r(u)}\,,\quad
r(u)={1-\sqrt u\over1+\sqrt u}\,,
\fe
as shown in Figure~\ref{fig:eyeframes}(a). In this frame the Hamiltonian is,
\ie
H_{\rm rad}(u)
={1+u\over2} D
+{1-u\over4}(P_0+K_0)\,,
\label{Hradialaffine}
\fe
which is Hermitian under the standard adjoint operation in radial quantization \cite{Simmons-Duffin:2016gjk}. Thus the variational argument is unchanged as expected.

For a fixed $u$, one may instead use the following radial coordinate
\ie
w_u={1+\sqrt u\,z\over1-\sqrt u\,z}\,.
\fe
This is the conventional map that sends the cusp points to $0$ and $\infty$, while the defect legs become straight rays
\ie
\arg w_u=\pm{\theta\over2}\,,
\fe
as in Figure~\ref{fig:eyeframes}(b). The two frames are related by
\ie
w_u={w+r(u)\over1+r(u)w}\,,
\fe
a M\"obius transformation preserving the unit circle generated by $P_0-K_0$.

Finally, writing
\ie
w_u=e^{\tau_u+i\varphi_u}
\fe
gives the cylinder frame of Figure~\ref{fig:eyeframes}(c), where $\Sigma_0$ lies at $\tau_u=0$ and the two eyelids become static lines at $\varphi_u=\pm\theta/2$. In this frame the cusp points sit at $\tau_u\to\pm\infty$, making manifest the standard interpretation of $\Gamma_{ab}(\theta)$ as a cylinder ground-state energy, while the aforementioned fixed radial frame makes the affine structure responsible for conformal concavity manifest. 
\section{No improvement from non-planar eye}
\label{app:nonplane}

In the main text we restricted the eye to a plane. One may ask whether more general conformal deformations, in particular those that bend the eye into transverse directions, lead to stronger constraints on $\Gamma_{ab}(\theta)$. We show here that they do not. Such global conformal deformations preserve the physical cusp angle and only change the conformal frame used for the variational problem. We first make this redundancy explicit for an asymmetric planar eye, which also provides a natural setting to clarify the conformal algebra in the presence of the defect.

The main text uses the familiar conformal algebra
\ie
[D,H]=P\,,
\quad
[D,P]=H\,,
\quad
[H,P]=-D\,.
\fe
We emphasize that these generators arise from the charges
\eqref{charge} evaluated on $\Sigma_0$ in the presence of the defect
punctures. Their definition therefore requires the appropriate local
renormalization at the punctures, reflecting the short-distance singularities
in the OPE of $T_{\mu\nu}$ with the defect \cite{Girault:2025kzt}.
Moreover, although the charge algebra is unchanged, its realization inside
defect correlation functions is nontrivial because the broken generators
$D$ and $P$ act nontrivially on the defect embedding
\cite{Gabai:2025zcs,Girault:2025kzt,Kong:2025sbk,Belton:2025ief,Drukker:2025dfm,Belton:2026xaw}.

By contrast, consider the conformal charges defined on a hypersurface
$\Sigma$ lying entirely away from the defects. Their algebra follows directly
from the usual stress-tensor OPE,
\ie
[\hat D,\hat{H}]
=
\hat{P}\,,
\quad
[\hat D,\hat{P}]
=
\hat{H}\,,
\quad
[\hat{H},\hat{P}]
=
-\hat D\,.
\label{ambalgebra}
\fe
Since the conformal current is conserved away from operator and defect
insertions, $\Sigma$ may be continuously deformed without changing the
charge. When $\Sigma$ is pushed towards $\Sigma_0$, however, it must wrap
around the semi-infinite portions of the line defects ending on $\Sigma_0$.
The defect Ward identity then expresses the resulting difference between
$\hat Q_\xi$ and the renormalized charge $Q_\xi$ on $\Sigma_0$ in terms of
integrated displacement operators, together with contributions localized at
the cusp endpoint.

To make this relation explicit, it is convenient to work in the cylinder
frame (see Figure~\ref{fig:eyeframes}(c)) and consider static line defects at
$\varphi=\varphi_i$. Deforming the charge surface as described above gives,
when acting on the cusp configuration,
\ie
\hat D
&=
D-\left[\int_{-\infty}^{0}d\tau\,\cosh\tau\,
\mathbb X(\tau)\right]_{\rm ren}
+\text{cusp terms}\,,
\\
\hat P
&=
P-\left[\int_{-\infty}^{0}d\tau\,\sinh\tau\,
\mathbb X(\tau)\right]_{\rm ren}
+\text{cusp terms}\,,
\fe
while $\hat H=H$ up to the corresponding localized terms at the cusp
\cite{Girault:2025kzt}. Here
\ie
\mathbb X(\tau)
\equiv
\sin\varphi_{1}\,\mathbb D^{a}(\tau)
+\sin\varphi_{2}\,\mathbb D^{b}(\tau)\,,
\label{Xdef}
\fe
where $\mathbb D^{a,b}$ denote the displacement operators in the $t$-$y$
plane on the line defects $a,b$, respectively.

At fixed defect embedding, the corresponding equal-time commutation relations
between the broken charges and the preserved charge $H$ on the quantization
slice are, 
\ie
[D,H]_{\rm fixed} 
&=
P-\mathbb X(0)\,,
\\
[P,H]_{\rm fixed} 
&=
D\,.
\label{fixedcommOS}
\fe
The commutator of the two broken charges, $[D,P]_{\rm fixed}$, is substantially more
subtle: it probes the nonlinear dependence on the defect shape, which in the
displacement-operator representation appears through contact terms generated
by repeated displacement insertions
\cite{Gabai:2025zcs,Girault:2025kzt,Kong:2025sbk,Belton:2025ief,Drukker:2025dfm,Belton:2026xaw}.

\subsection{Asymmetric planar eye}

Let us now specialize to the configuration describing an asymmetric eye, which in the cylinder frame has eyelids located at
\ie
\varphi_1=-\theta_-\,,
\quad
\varphi_2=\theta_+\,,
\quad
\theta_-+\theta_+=\theta\,,
\fe
and we define
\ie
\delta s\equiv\sin\theta_-+\sin\theta_+\,.
\fe
The displacement Ward identity, together with translation symmetry in $\tau$, gives expectation values taken in the asymmetric half-eye state
\ie
\la\mathbb D^a\ra
=
-\la\mathbb D^b\ra
=
\Gamma'_{ab}(\theta)\,,
\quad
\la\mathbb X\ra
=
-\delta s\,\Gamma'_{ab}(\theta)\,.
\fe
Taking the diagonal matrix element of the first relation in \eqref{fixedcommOS} in the $H$ ground state therefore gives
\ie
\la P\ra
=
-\delta s\,\Gamma'_{ab}(\theta)\,.
\label{Pmoment}
\fe
Consequently the affine family of Hamiltonians introduced in the main text has expectation value
\ie
\la{H}(u)\ra
=
{1+u\over2}\Gamma_{ab}(\theta)
-{1-u\over2}\,\delta s\,\Gamma'_{ab}(\theta)\,.
\label{asymmoment}
\fe
The limit $u=0$ is particularly simple, since
$H(0)=P_0/2$. For parallel conformal defects separated by a transverse distance $L$ in flat space, the ground state energy takes the form
\ie
E_{ab}(L)
=
-{\varepsilon_{ab}\over L}
\quad
\varepsilon_{ab}\equiv -C_{ab\bar c}\,,
\fe
where $c$ labels the dominant fusion channel. We thus have
\ie
\Gamma_{ab}(\theta)
-\delta s\,\Gamma'_{ab}(\theta)
\geq
-{\varepsilon_{ab}\over\sqrt{u_+}+\sqrt{u_-}}\,,
\label{asymeye}
\fe
where
\ie
u_\pm
=
\tan^2{\theta_\pm\over2}\,.
\fe
Applying the same argument upon inversion, where the Hamiltonian is proportional to $-K_0$, gives instead
\ie
\Gamma_{ab}(\theta)
+\delta s\,\Gamma'_{ab}(\theta)
\geq
-{\varepsilon_{ab}\over1/\sqrt{u_+}+1/\sqrt{u_-}}\,.
\fe
The one-sided limit
\ie
(\theta_-,\theta_+)\to(0,\theta)
\fe
then gives
\begin{align}
\Gamma_{ab}(\theta)
-\sin\theta\,\Gamma'_{ab}(\theta)
&\geq
-\varepsilon_{ab}\cot{\theta\over2} \\ 
\Gamma_{ab}(\theta)
+\sin\theta\,\Gamma'_{ab} (\theta)
&\geq 0\,.
\label{onesidedasym}
\end{align}
For a general junction the second inequality is weaker than the smooth limit daughter inequality derived in the main text \eqref{gamma1d}
and agrees with it when $\Delta_{ab}=0$.

\subsection{Non-planar eye}

We now ask whether non-planar eye geometries can produce stronger inequalities, which are sensitive to higher dimensions. The analysis can be performed entirely on the $t=0$ spatial slice $\Sigma_0 \simeq \mathbb R^{d-1}$,
or equivalently in the radial or cylinder frames. 

Let $\vec x\in\mR^{d-1}$ denote coordinates on $\Sigma_0$, and choose a unit vector $\hat e$ along the spatial direction of the planar eye. Its two punctures are located at
\ie
\vec x_{1,2}
=
\pm r\,\hat e\,,
\quad
r=\tan{\theta\over4}\,.
\label{symmpunctures}
\fe
More generally, two punctures $\vec x_1,\vec x_2$ determine an eye  with opening angle $\theta$ through
\ie
\sin{\theta\over2}
=
{|\vec x_1-\vec x_2|
\over
\sqrt{(1+|\vec x_1|^2)(1+|\vec x_2|^2)}}\,.
\label{cutangle}
\fe

Restricting to conformal transformations that preserve the NS quantization
slice $\Sigma_0$, and quotienting by the stabilizer of $H$ within this subgroup, a
general deformation of the NS frame (changing the locations of $N$ and $S$ as well as the punctures on $\Sigma_0$) is generated by
\ie
G_{\chi,\vec\beta}
\equiv
\chi D+{K_{\vec\beta}-P_{\vec\beta}\over2}\,,
\quad
P_{\vec\beta}\equiv\beta^iP_i\,,
\quad
K_{\vec\beta}\equiv\beta^iK_i\,.
\label{Gchibeta}
\fe
Its adjoint action on the conformal Hamiltonian defines
\ie
[G_{\chi,\vec\beta},H]
=
V_{\chi,\vec\beta}
\equiv
\chi P+\beta^iM_{0i}\,.
\fe
Choosing $\chi^2+|\vec\beta|^2=1$, the conformal algebra further gives
\ie
[G_{\chi,\vec\beta},V_{\chi,\vec\beta}]
=
H\,,
\fe
and hence
\ie
e^{-\eta G_{\chi,\vec\beta}}
H
e^{\eta G_{\chi,\vec\beta}}
=
H\cosh\eta
-
V_{\chi,\vec\beta}\sinh\eta\,.
\fe
Removing the overall factor $\cosh\eta$ and defining
$\lambda=\tanh\eta$ therefore produces the affine family
\ie
H_{\chi,\vec\beta}(\lambda)
\equiv
H-\lambda V_{\chi,\vec\beta}\,.
\label{generalbendH}
\fe

Choosing $\chi^2+|\vec\beta|^2=1$, the conformal algebra gives
\ie
[G_{\chi,\vec\beta}, V_{\chi,\vec\beta}]
=
 H\,,
\fe
and hence
\ie
e^{-\eta G_{\chi,\vec\beta}}
H
e^{\eta G_{\chi,\vec\beta}}
=
H\cosh\eta
- V_{\chi,\vec\beta}\sinh\eta\,.
\fe
With $\lambda=\tanh\eta$, this becomes the affine family
\ie
H_{\chi,\vec\beta}(\lambda)
\equiv
H-\lambda V_{\chi,\vec\beta}\,.
\label{generalbendH}
\fe

On $\Sigma_0$, both $H$ and $V_{\chi,\vec\beta}$ have vanishing tangential components, so the defect punctures remain fixed and \eqref{generalbendH} acts on the same two-defect Hilbert space $\cH_{ab}$. Since $V_{\chi,\vec\beta}$ is self-adjoint, this is an affine family of self-adjoint Hamiltonians, exactly as in the variational argument of the main text. 

Restricting $G_{\chi,\vec\beta}$ to $\Sigma_0$,  the corresponding conformal Killing vector is
\ie
v_{\chi,\vec\beta}(\vec x)
=
\chi\vec x
+(\vec\beta\cdot\vec x)\vec x
-{1+|\vec x|^2\over2}\vec\beta\,.
\label{cutbendCKV}
\fe
Let $g_\eta$ denote its finite flow and define
\ie
\vec x_i(\lambda)
\equiv
g_\eta(\vec x_i)\,,
\quad
\lambda=\tanh\eta\,.
\fe
We denote by $\theta_\lambda$ the opening angle of the $H$-static eye through these two punctures. By \eqref{cutangle},
\ie
\sin{\theta_\lambda\over2}
=
{|\vec x_1(\lambda)-\vec x_2(\lambda)|
\over
\sqrt{
(1+|\vec x_1(\lambda)|^2)
(1+|\vec x_2(\lambda)|^2)
}}\,,
\quad
\theta_0=\theta\,.
\label{globalbendangle}
\fe
Note the distinction between $\theta$ and $\theta_\lambda$. A conformally transformed copy of the original eye still has local cusp angle $\theta$. The angle $\theta_\lambda$ instead labels the eye for the defect configuration where $H$ is a conserved charge, through the transformed punctures, which is the conformally related problem used to evaluate the spectrum of $ H_{\chi,\vec\beta}(\lambda)$.\par
Conformal covariance therefore identifies its lowest energy as
\ie
\cE_{\chi,\vec\beta}(\lambda)
=
\sqrt{1-\lambda^2}\,
\Gamma_{ab}(\theta_\lambda)\,.
\label{globalbendenergy}
\fe
Since $H_{\chi,\vec\beta}(\lambda)$ is affine in $\lambda$, the variational principle implies concavity for the ground state energy. Wherever the lowest branch is twice differentiable, we have
\ie
-\left.
{d^2\cE_{\chi,\vec\beta}\over d\lambda^2}
\right|_{\lambda=0}
\geq0\,.
\label{genconcav}
\fe

To resolve the resulting Hessian, we decompose
\ie
\vec\beta
=
\beta_\parallel\hat e
+\vec\beta_\perp\,,
\quad
\vec\beta_\perp\cdot\hat e=0\,.
\fe
The scalar Hessian has three symmetry channels: the dilation direction $\chi$, the longitudinal direction $\beta_\parallel$, and the transverse subspace $\vec\beta_\perp$, whose multiplicity is $d-2$.

Evaluating \eqref{cutbendCKV} at the punctures \eqref{symmpunctures} gives
\ie
\left.
{d\vec x_{1,2}(\lambda)\over d\lambda}
\right|_{\lambda=0}
=
\pm\chi r\,\hat e
-{1-r^2\over2}\beta_\parallel\hat e
-{1+r^2\over2}\vec\beta_\perp\,.
\label{puncturemotions}
\fe
Thus $\chi$ moves the punctures oppositely along their common axis and changes their separation, whereas $\beta_\parallel$ moves them together along that axis and $\vec\beta_\perp$ moves them together transverse to the original eye plane. Crucially, only the dilation changes the opening angle at linear order. Indeed, differentiating \eqref{globalbendangle} gives
\ie
&\left.
{d\theta_\lambda\over d\lambda}
\right|_{\lambda=0}
=
2\chi\sin{\theta\over2}\,,
\\
&
\left.
{d^2\theta_\lambda\over d\lambda^2}
\right|_{\lambda=0}
=
-2\tan{\theta\over2}
\left[
1-
\left(1+\cos^2{\theta\over2}\right)\chi^2
-\sin^2{\theta\over2}\,\beta_\parallel^2
\right]\,.
\label{globalbendexp}
\fe
The concavity \eqref{genconcav} can therefore be written as
\ie
-\left.
{d^2\cE_{\chi,\vec\beta}\over d\lambda^2}
\right|_{\lambda=0}
=
\chi^2M_{\rm dil}
+\beta_\parallel^2M_\parallel
+|\vec\beta_\perp|^2M_\perp
\geq0\,,
\label{fullhessian}
\fe
with
\ie
M_{\rm dil}
=
\Gamma_{ab}
-\sin\theta\,\Gamma'_{ab}
-4\sin^2{\theta\over2}\Gamma''_{ab}\,,
\label{Mdil}
\fe
and
\ie
M_\parallel
=
\Gamma_{ab}
+\sin\theta\,\Gamma'_{ab}\,,
\quad
M_\perp
=
\Gamma_{ab}
+2\tan{\theta\over2}\Gamma'_{ab}\,.
\label{hessianchannels}
\fe

The dilation channel $M_{\rm dil}$ is precisely the mother inequality \eqref{gamma2d}. It is also the only channel sensitive to $\Gamma''_{ab}$, since only $\chi$ changes $\theta_\lambda$ at first order. The longitudinal channel is already controlled more strongly by \eqref{gamma1d},
\ie
M_\parallel
=
\Gamma_{ab}
+\sin\theta\,\Gamma'_{ab}
\geq
\Delta_{ab}\sin{\theta\over2}
\geq0\,.
\label{Mparallelbound}
\fe
The transverse Hessian condition is also weaker. Define
\ie
\bar{\cE}_{ab}(u)
\equiv
\cE_{ab}(u)
-{\Delta_{ab}\over2}(1+u)\,.
\label{Ebarsub}
\fe
Since the subtraction is affine,
\ie
\bar{\cE}_{ab}''(u)
=
\cE_{ab}''(u)
\leq0\,.
\fe
Using the smooth limit data
\ie
\bar{\cE}_{ab}(1)=0\,,
\quad
\bar{\cE}'_{ab}(1)=-A_{ab}\,,
\label{Ebarendpoint}
\fe
the transverse Hessian can be written as
\ie
M_\perp
=
{2\sqrt{u}\over1-u}\,X_\perp(u)\,,
\quad
X_\perp(u)
\equiv
(1+u)\bar{\cE}'_{ab}(u)
-\bar{\cE}_{ab}(u)\,.
\label{MperpE}
\fe
Conformal concavity gives
\ie
X_\perp'(u)
=
(1+u)\bar{\cE}_{ab}''(u)
\leq0\,,
\quad
X_\perp(1)=-2A_{ab}\,,
\fe
and hence, for $0<u<1$,
\ie
M_\perp
\geq
-{4\sqrt{u}\over1-u}\,A_{ab}\,.
\label{Mperpbound}
\fe
For a nondegenerate DCO that is a scalar under transverse rotations, $A_{ab}=0$. More generally, as explained around \eqref{eq:boundonDCO}, with suitable transverse reflection symmetries,
\ie
A_{ab}\leq0\,.
\fe
Combining this with \eqref{Mperpbound} gives
\ie
M_\perp
\geq
-{4\sqrt{u}\over1-u}\,A_{ab}
\geq0\,.
\fe
Thus the transverse Hessian condition $M_\perp\geq0$ is already implied by planar conformal concavity, and is strictly weaker whenever $A_{ab}<0$. For $A_{ab}=0$ the two bounds coincide. Note that in $d=2$ the transverse sector is absent altogether.

To summarize, an arbitrary conformal deformation of the NS transfer frame for the planar eye does not lead to an improved inequality for $\Gamma_{ab}(\theta)$. The dilation direction is the unique channel probing $\Gamma''_{ab}$ and reproduces the mother inequality, while the longitudinal and transverse channels are weaker consequences of the same conformal concavity and its endpoint data.

\bibliography{bibliography}
\end{document}